\documentclass[12pt]{article}
\usepackage[english]{babel}
\usepackage{amssymb,graphicx,wrapfig}
\author{Vladimir A. Petrov\footnote{Vladimir.Petrov@ihep.ru} 
 and Nikolai P. Tkachenko\footnote{Nikolai.Tkachenko@ihep.ru}}
\index{}
\title{ATLAS vs TOTEM: Disturbing Divergence.}
\date{}

\begin{document}

\maketitle
\begin{center}

A.A. Logunov Institute for High Energy Physics 

NRC "Kurchatov Institute", Protvino, RF
\end{center}

\begin{abstract}
With use of a new $ C+N $ amplitude which correctly accounts for CNI to all
orders in $\alpha$ the values of the parameters $\rho^{pp}$,
$\sigma_{\mbox{tot}}^{pp}$ and $B^{pp}$ are analysed on the basis of data
published by the TOTEM and ATLAS collaborations at energy $\sqrt{s}= 13$
TeV . The result of both joint and separate analyses of these data confirmed,
in particular, a  noticeable difference between the central values of
$\sigma^{pp}_{tot}$, as measured by the TOTEM and by the ATLAS-ALFA, reaching
up to 8 standard deviations ATLAS experiment.
 
\end{abstract}

\section*{Introduction}
On a wide palette of experiments at the LHC, those dealing with diffraction
scattering occupy a relatively modest place. There are actually two of
them, viz., the TOTEM collaboration and the ALFA group in the ATLAS collaboration.
If we take the TeV energy region in general, then the precursors of such
measurements were those that were taken at the Tevatron $\bar{p}p$ collisions
about 2 TeV at CDF, E-710 and E-811 \cite{T}.

The paucity of these experiments does not in any way imply their secondary
importance. After the discovery of the Higgs boson, one of the main problems
of the Standard Model is the problem of  behaviour of color fields at large
distances, the notorious "confinement problem of QCD". The kinematic region
of the noted diffraction experiments is a peculiar interplay of the ultraviolet
region (large collision energies) and the infrared region (very small momentum
transfers) i.e. just the area where the effects of confinement are
dominant\footnote{We note that high energies do not at all make it possible
to use the advantages of "asymptotic freedom" (anti screening at small distances),
since the average transverse distances in the experiments under discussion
exceed 1 fermi and \textit{grow} with increasing energy.}.

Thus the conceptual value and problematic level of diffraction measurements
are beyond doubt.Therefore, it is so important for theorists and phenomenologists
to have reliable data from these measurements.However, already the data published
by the three above-mentioned collaborations at the Tevatron in the period
1992-94 \cite{T} led and still lead to some confusion. Indeed,while the
data on $\sigma_{tot}^{\bar{p}p}$ from E710 and E811 were more or less consistent
within the error bars, the CDF data  turned out to be as far as more than
3 standard deviations from those two experiments. Unfortunately, the participants
in these three experiments were unable (or unwilling) to analyse that unacceptable
situation, preferring to leave the community completely bewildered.  Some
hopes for a clarification of the situation have eventually dissipated and
turned into an unfortunate but inevitable circumstance.

The launch of the LHC, with TOTEM and ALFA/ATLAS, produced new hopes
for new data on diffraction scattering in a new range of energies. However
at energies of 7 and 8 TeV, a gap between the values of $\sigma_{tot}^{pp}$
measured by the two collaborations  arose and
not only did not decrease, but even increased and reached its maximum at
13 TeV (see below)).

In our recent paper \cite{PRD106} we studied the extraction of the "forward"
parameters $\rho^{pp}, \sigma_{tot}^{pp}$ and $B^{pp}$ from the TOTEM
experimental data \cite{TOTEM} at energy $\sqrt{s} = 13$ TeV based on
amplitudes taking into account CNI up to $\mathcal{O} (\alpha^{2})$.
We managed to show that the values of $ \rho $, claimed by the TOTEM group
as retrieved from their experimental data, cannot be determined with that
high precision that the authors of the experiment point to. This showed the
need to carefully check the published data.

Since then, a new information has become available, viz., the data on diffraction
scattering from the ATLAS experiment at $13$ TeV, which are posted in HEP
DATA \cite{ATLAS} and  which have shown that these new data on
three basic "forward" observables ( mostly on $\sigma_{tot}$) at
$\sqrt{s}= 13$ TeV only increased even more the discrepancy between the two
experiments, already observed at $\sqrt{s}= 8$ TeV.

If this situation is repeated (or even aggravated) at the last run at
14 TeV, then, say, the clarification of the Odderon question may
"hang" for many years, thereby devaluing many of the efforts of both groups
during the last decade.

Desire to better understand the source of the said discrepancy prompted us
to carry out a new, comparative analysis of the sets
of basic parameters from both experiments, especially since this time we
have the opportunity to apply a more accurate $C+N$ scattering amplitude
that takes into account the CNI to all orders in $\alpha$.
 
\section{Experimental  Data from ATLAS and TOTEM}

\begin{figure}[htb]
\noindent
$$
\includegraphics[width=85mm]{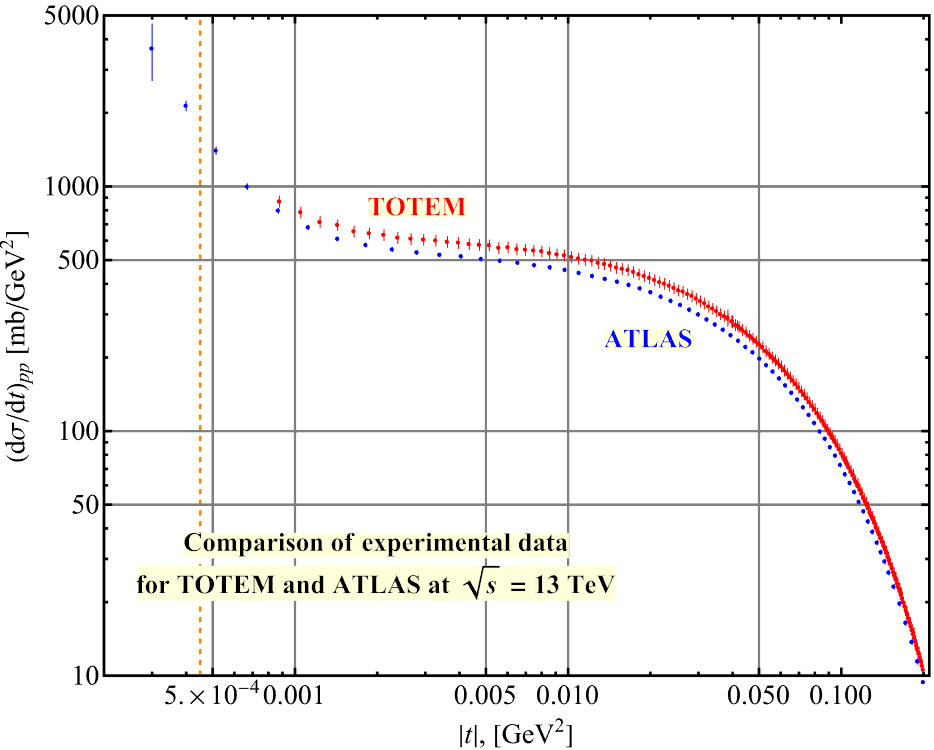}
~
\includegraphics[width=85mm]{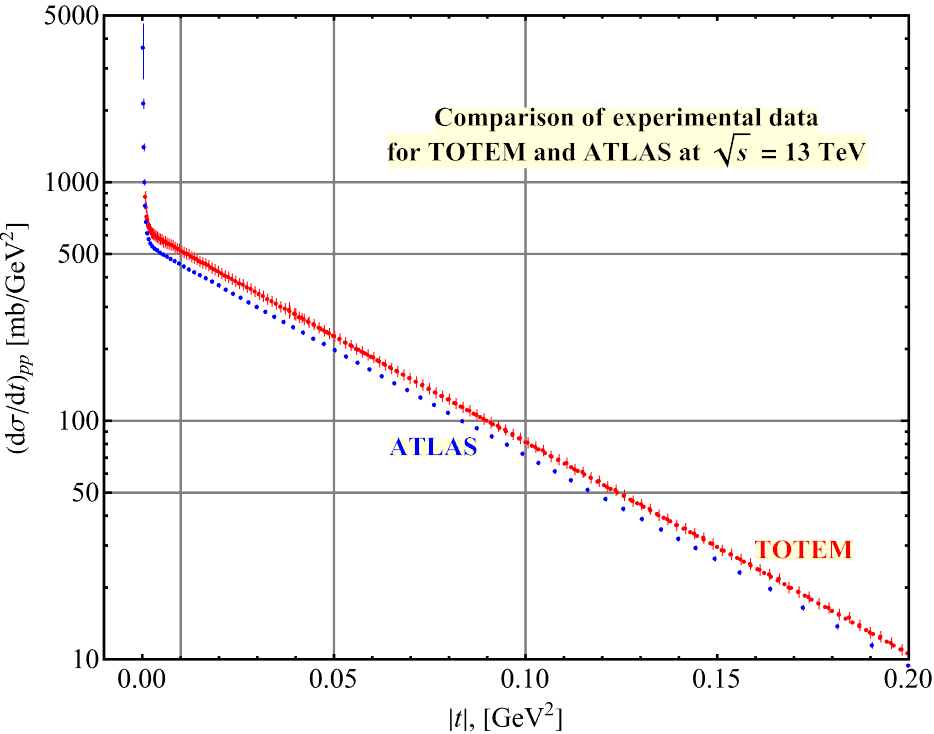}
$$
\caption{{\it Differential cross sections $pp$ at $\sqrt{s} = 13$
TeV measured by ATLAS and TOTEM in double logarithmic (left) and
semilogarithmic scales.}}
\label{Pic1}
\end{figure}

Fig. \ref{Pic1} shows $d{\sigma _{pp}}/dt$ measured at the ATLAS and TOTEM
facilities at $\sqrt{s} = 13$ TeV. It is clearly seen that they differ significantly
from each other even taking into account large errors at the TOTEM facility:
the ATLAS  data systematically lie below the TOTEM  data. A priori, it is
not even clear how correct the operation of joint fitting of these data would
be. Nevertheless, we are going to carry out such an operation and to see
the results.It is clear in advance that these results should be treated only
as estimates. According to the recommendations of the authors of ATLAS, we
do not include two experimental points with minimal values of $|t|$ in the
analysis,although we keep these two points on the graphs. Thus, only the
experimental points for which $|t|>0.00045~\mbox{GeV}^2$ are used further
in all fits (dashed vertical line on the left graph of Fig. \ref{Pic1}).

\newpage

\section{Model description}

We follow the model described in \cite{PRD106}. However, unlike \cite{PRD106},
we will use the $ C+N $ amplitude which takes into account all order in
$\alpha$. The main parametrisations and other ingredients  are as follows:
\begin{equation}
T_{N}(q^2)=T_{N}(0,s)e^{-Bq^{2}/2}=
\frac{s(i+\rho)\sigma_{\mbox{tot}}}{(\hbar c)^{2}}
e^{-Bq^{2}/2};
\label{eq1}
\end{equation}
this is the pure strong interaction ("nuclear") scattering amplitude\footnote{We
are aware that this very simplified parametrization is far from perfect.
For example, it does not take into account the dependence of the phase on
the momentum transfer, which, strictly speaking, is unacceptable \cite{Pe'},
etc. However, as an approximation in the field of relevant transfers, it
is quite acceptable. What is most important, we \textit{must} use model (1),
because it is actually used by both collaborations, TOTEM and ATLAS, for
data processing. Otherwise, comparing our parameters with theirs would be
inconsistent.} and $(\hbar c)^{2}=0.389379$ [mb$\cdot \mbox{GeV}^2$].

The formula for the Coulomb-nuclear scattering amplitude was derived in \cite{Pe}:
\begin{equation}
T_{C+N}=-2is\Xi(q^2)+\frac{1}{4\pi}\int_{0}^{\infty}dq'^2
\Xi(q'^{2}){\bar T}_{N}(q'^2,q^2),
\label{eq3}
\end{equation}
where \footnote{$J_{0}(z)$ is a Bessel function of the first kind \cite{Grad}.}
\begin{equation}
\Xi(q)=2\pi\int_{0}^{\infty}db~b~J_{0}(qb)e^{2i\delta_{c}(b)},
\label{eq4}
\end{equation}
and  $\delta_{c}(b)$ is expressed in terms of the Meijer
G-function\footnote{Meijer G-function is a generalized hypergeometric
function \cite{Grad} defined in terms of the Mellin-Barnes integral,
and the integral in(\ref{eq5}) is calculated via the Mellin
transform.
$$
\mbox{G}(x)\equiv
\mbox{G}
\left[
\left\{ \{ a_1,...,a_n \},\{ a_{n+1},...,a_{p} \} \right \} ,
\left\{ \{ b_1,...,b_m \},\{ b_{m+1},...,b_{q} \} \right \} , x\right] =
\frac{1}{2\pi i}
\int
\frac{
\prod_{j=1}^{m}(b_{j}-s) \prod_{j=1}^{n}(1-a_{j}+s)
}{\prod_{j=m+1}^{q}(1-b_{j}+s) \prod_{j=n+1}^{p}(a_{j}-s)}x^{s}
ds\vspace{-3.1mm}
$$}
\begin{equation}
\delta_{c}(b)=\alpha\int_{0}^{\infty}(dk/k)F^{2}(k^2)[1-J_{0}(kb)]=
\frac{\alpha}{12} \mbox{G}\left[
\left\{ \{ 1,1 \},\{~\}\right \} ,
\left\{ \{ 1,4 \},\{ 0,0 \}\right \} ,
\left( \frac{b\Lambda}{2} \right)^{2}
\right] ,
\label{eq5}
\end{equation}.

\begin{equation}
{\bar T}_{N}(q'^{2},q^2)\equiv T_{N}(q^2)e^{-Bq'^{2}/2}I_{0}(Bq'q)=
\frac{s(i+\rho)\sigma_{\mbox{tot}}}{(\hbar c)^{2}}
e^{-B(q'^{2}+q^2)/2}I_{0}(Bq'q),
\label{eq2}
\end{equation}
where $I_{0}(x)$ is a modified Bessel function of the first kind (Infeld
function) \cite{Grad}. We use, as in \cite{PRD106},  the dipole electromagnetic
form factor:
$$
F(q^2) = (1+q^2/\Lambda)^{-2},
$$
where $\Lambda=\sqrt{0.71}$ [GeV].
Note that for $ \alpha  $ we use $ \alpha=1/137 $ exactly instead of
$1/137,036047... $. The difference is negligible within the accuracy used
by us\footnote{We also neglect the scale dependence of $ \alpha $ which,
as shown in \cite{Ca}, is also negligible.}. 

\begin{wrapfigure}[18]{l}{121mm}
\vspace{-0.1mm}
\includegraphics[width=122mm]{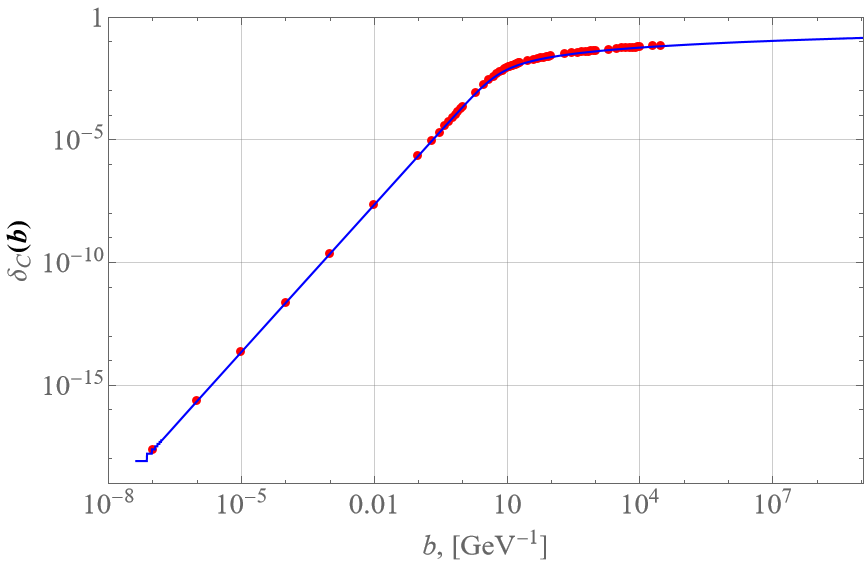}
\vspace{-7.1mm}
\caption{{\it Fitting curve (blue line) and exactly calculated values of
the $\delta_{c}(b)$ function, which are indicated by red dots.}}
\label{Pic2}
\end{wrapfigure}

By direct substitution (\ref{eq5}) into (\ref{eq4}), the integral is not
calculable analytically. However, the function $\delta_{c}(b)$ allows a good
approximation:
\begin{equation}
\delta_{c}^{appr}(b)\simeq\frac{\alpha}{2}\cdot
\mbox{ln} \left[ 1+\left(  \lambda b \right)^2 \right],
~~~\lambda \equiv \frac{3\Lambda}{10} = 0.2528...\mbox{~[GeV]}, 
\label{eq6}
\end{equation}
the graph of which we present in Fig. \ref{Pic2}
\newpage

With such an approximation, the integral in the expression (\ref{eq4}) is
calculated analytically and the function $\Xi(q)$ becomes\footnote{For reference:
$\int_{0}^{\infty}\Xi(q')dq' = 4\pi$, i.e. the integral converges, which
implies the convergence of the integral in the final expression for the total
amplitude (\ref{eq3})}:
\begin{equation}
\Xi(q)=\frac{2\pi}{\lambda^{2}\Gamma\left( -i\alpha \right)}\cdot
\frac{K_{1+i\alpha}\left(  q / \lambda\right)}
{\left( q/2\lambda \right)^{1+i\alpha}}
\label{eq7}
\end{equation}
where $K_{\mu}(z)$ is a modified Bessel function of the second kind (McDonald
function) \cite{Grad}. The function $ \Xi(q) $ has dimension [$\mbox{GeV}^{-2}$].
Accordingly, the total amplitude $T_{C+N}$ turns out to be dimensionless.

Substituting this expression into (\ref{eq3}) again leads to an integral
that is not taken analytically. However, if the function $I_{0}(z)$ is expanded
in a Taylor series in a neighbourhood of zero,which dominates the integral
in (3), then the integral can be  calculated analytically term by term.
As we do not use data for which $|t|> 0.2$ then direct calculations show
that six significant figures in this integral are completely provided by
the first eight terms of the mentioned expansion.

With account of all that, the differential cross section is calculated using
the standard formula:
\begin{equation}
\frac{d\sigma}{dt} = (\hbar c)^2 \frac{\left| T_{C+N}\right|^{2}}{16\pi s^2}
~~~ \left[ \frac{\mbox{mb}}{\mbox{GeV}^2} \right] .
\label{eq8}
\end{equation}

Further, everywhere the total expression for $\chi^2$ is compiled using
exactly this expression for the differential cross section using weight matrices.
Weight matrices of statistical and systematic errors of the ATLAS experiment
are listed in the HEP Data \cite{ATLAS}. Unfortunately,the TOTEM results
are (still?) not placed in the HEP Data. So, the corresponding matrix of
systematic errors in this experiment was just calculated from the tables
containing systematic errors given in \cite{TOTEM}. Since the TOTEM does
not give such a matrix, there is nothing left but to take it, using \cite{TOTEM},
in a diagonal form.
\newpage

\section{Extraction of parameters from ATLAS and TOTEM experiments}.
\vspace{-13.6mm}
\subsection{General remarks on the processing technique}

\begin{wrapfigure}[43]{l}{103mm}
\vspace{-5.1mm}
\includegraphics[width=103mm]{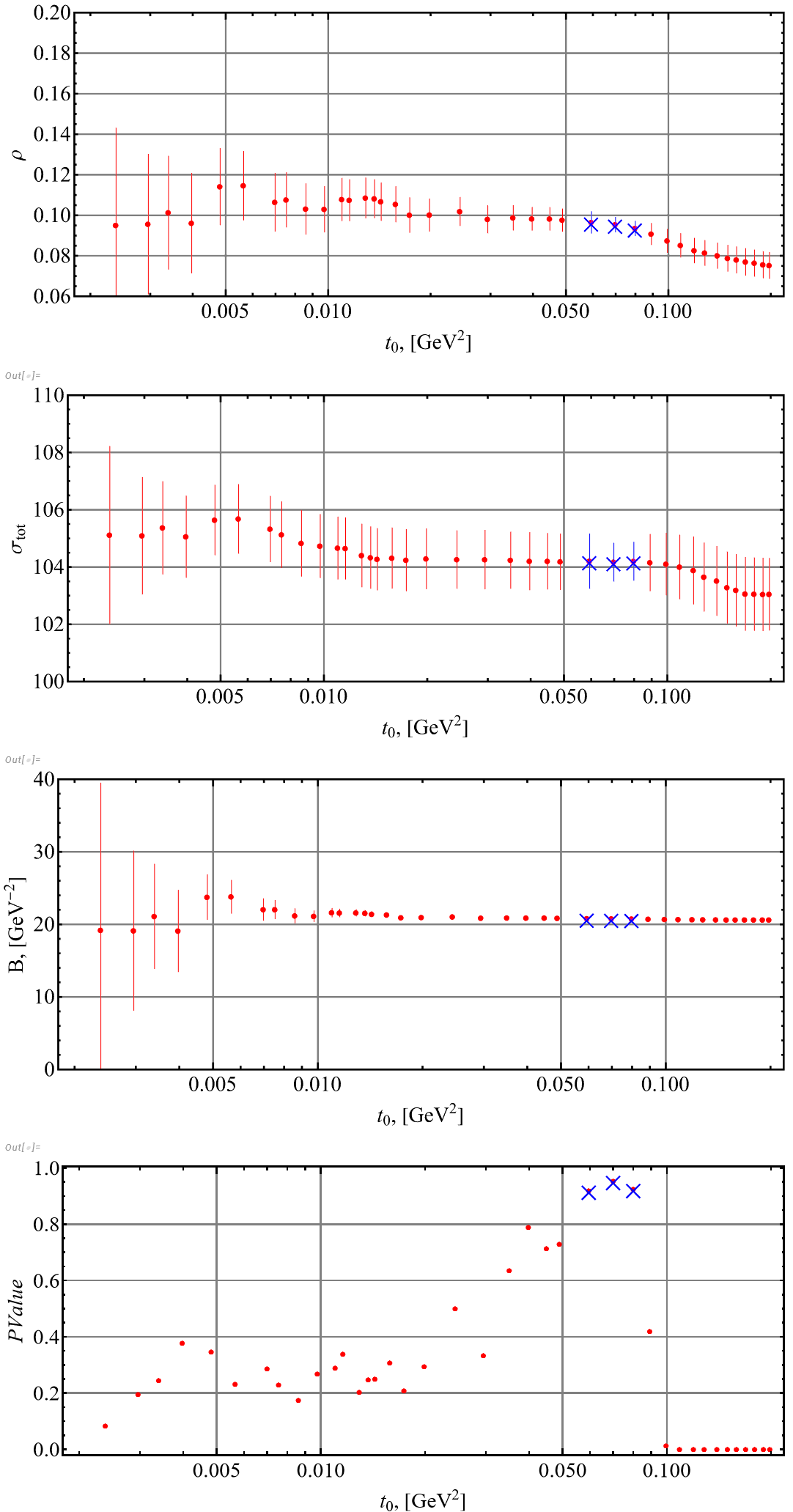}
\vspace{-6.6mm}
\caption{\textit{Behaviour of Model Parameters with Gradual Dropping
experimental data with large values of $|t|$, more than some
$t_{0}$ values. The crosses indicate the results for which the level
confidence above 90\%.}}
\label{Pic3}
\end{wrapfigure}

In most cases, as data with large $|t|$ is cut off, the extractable
parameters began to show extreme instability of their values. In that
case, we also discarded data with small values of $|t|$ at which the
instability of the behaviour of parameters takes place and the fit was carried
out on a data set in which the values of the transmitted momentum were limited
both from below and from above.
In all cases, we always indicate the region $|t|$ used for the
retrieval of parameters.

All fittings according to the above model were built on the basis of calculating
the function $\chi^2$ using the weight matrices obtained from the corresponding
covariance matrices.

Covariance matrices of systematic and statistical errors of the experiment
ATLAS were taken from the HEP Data \cite{ATLAS} database.

The TOTEM experiment did not post the results in HEP Data. However, in his
article \cite{TOTEM} they cited sources of systematic errors, according to
which we and built a covariance matrix of systematic errors in this experiment.
Data on the sources of statistical errors are absent in \cite{TOTEM} where
only total statistical errors are shown. So there was nothing left
but to use the covariance matrix of statistical errors in the diagonal
form.\vspace{-6.6mm}

\subsection{Joint fitting of two experiments}

The full covariance matrices were defined as the sum of the covariance matrices
of statistical and systematic errors and then the weight matrices were calculated.
With joint fitting, the covariance matrix was compiled as a block matrix
from the covariance matrices of two experiments. The rest of the elements
were assumed to be equal. zero, which corresponds to the absence of correlations
between the measurements in the two experiments. Such an assumption is hardly
admissible, since both facilities use the same beams. This is definitely
a weak point.
such an approach, which once again indicates that the results of the joint
considerations should be treated as estimates.
Fig. \ref{Pic3} shows the results of the joint fitting of the ATLAS and TOTEM
data, by the method of gradual cutting off the data with large values of
$|t|$. There are only three points for which P-value $ >0.9$ and they are
located side by side in the interval $0.06\lesssim t_{0}\lesssim 0.08~\mbox{GeV}^2$.
For larger values of $t_{0}$ the level of confidence is generally zero. In
the table below we present parameter values corresponding to this
cut-off\footnote{The minimum number of decimal places in the values of the
matrix and parameters is described in detail in \cite{Ezhela}}.

Based on Fig. \ref{Pic3} we see no reason to cut off any points with small
values of $|t|$, because taking into account
errors of parameter values at $t_{0}\lesssim 0.08~\mbox{GeV}^2$ lie
on one plateau. However, discarding a small number of such points makes
sense in order to see how they affect the stability of values
retrieved parameters. We therefore discarded the experimental points
for which $|t|<0.004~\mbox{GeV}^2$, as well as the points corresponding to
the condition $|t|>0.014~\mbox{GeV}^2$ i.e. we are considering experimental
points that satisfy the condition $0.004 <|t|<0.014~\mbox{GeV}^2$.
This corresponds approximately to the intervals we used to extract
parameters in \cite{PRD106}. At the same time, a high level of confidence
is maintained\footnote{Next, we will justify in more detail the reason for
discarding the experimental values with the smallest momentum transfers.}.
The results are as follows:\vspace{-3.1mm}
\begin{figure}[htb]
$$\includegraphics[width=130mm]{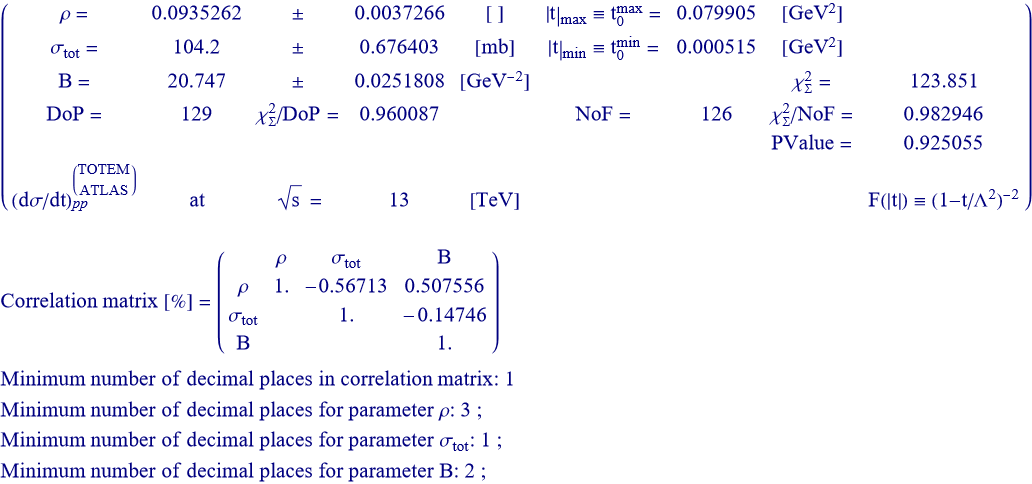}\vspace{-3.1mm}$$\label{Table1a}
$$\includegraphics[width=130mm]{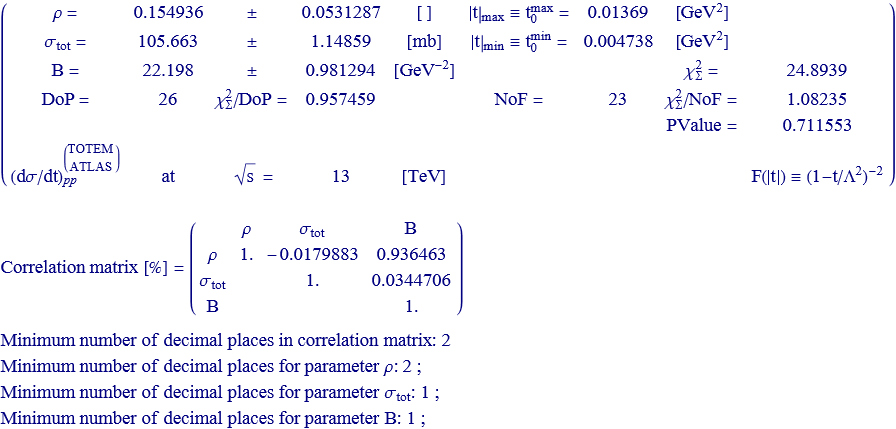}\vspace{-3.1mm}$$\label{Table1b}
\end{figure}

~\vspace{-11.1mm}

We see that there is a sharp change in the central values of the parameters
and a substantial increase in errors.

\begin{figure}[htb]
\noindent
$$
\includegraphics[width=95mm]{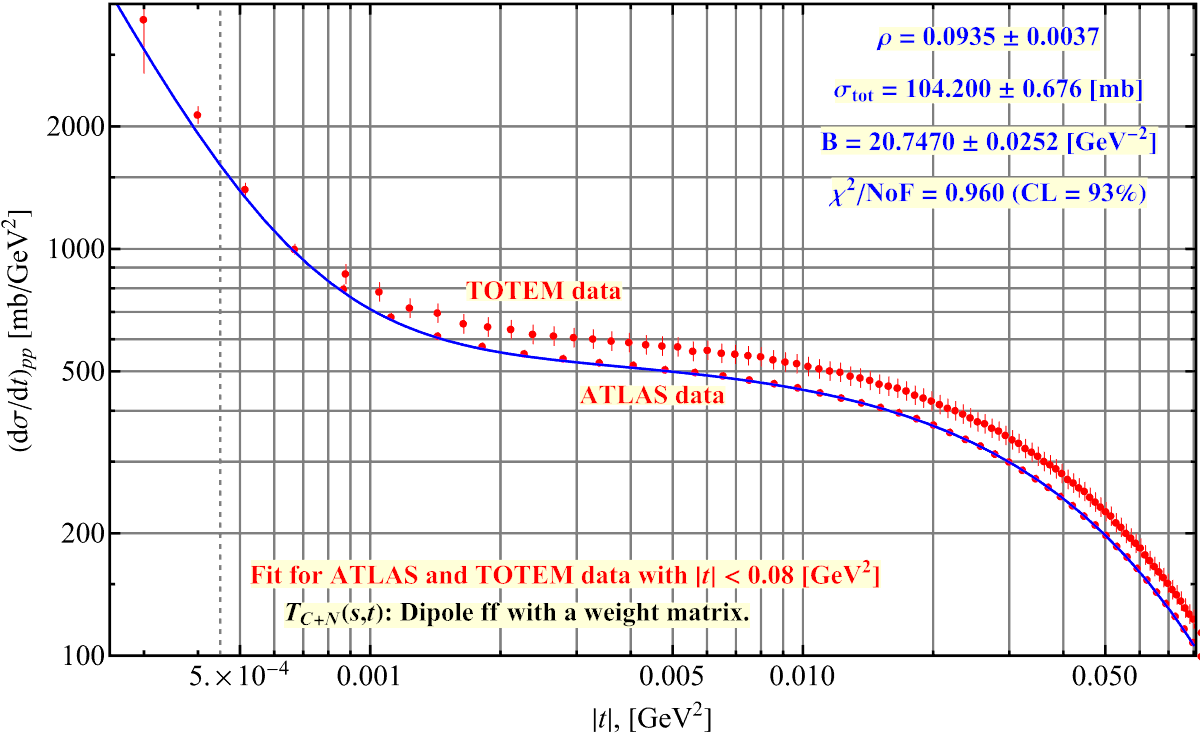}
~
\includegraphics[width=95mm]{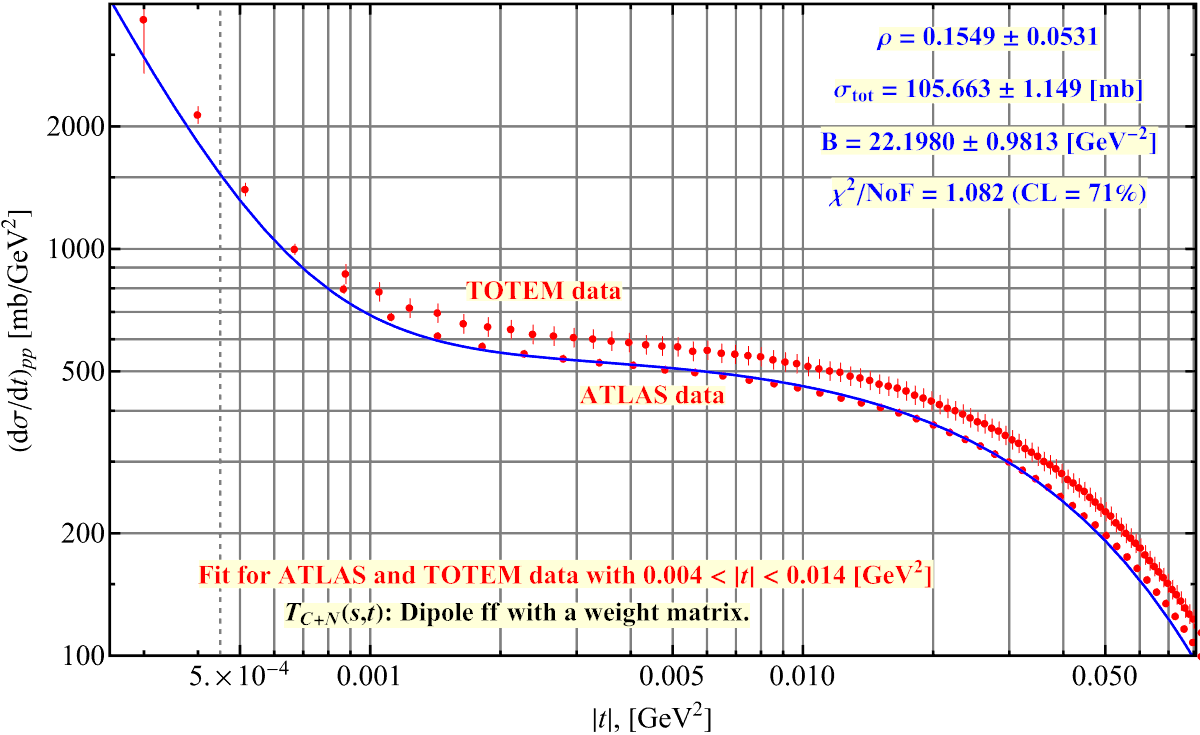} \vspace{-3.6mm}
$$
\parbox{190mm}{\caption{{\it Experimental data from ATLAS and TOTEM and
theoretical differential cross section curves corresponding to two
tables above. \vspace{-9.1mm}}}
\label{Pic4}}
\end{figure}

Fig. \ref{Pic4} shows graphs of theoretical curves corresponding to these
two cases.

The curves unambiguously gravitate towards the ATLAS experimental data, which
generally speaking, it is not surprising, because the experimental errors
in ATLAS significantly lower than the experimental errors of TOTEM.

However, as we said above, these results should be treated exclusively
as preliminarily estimated values due to significantly different experimental
measurements of the same process.
Some voluntarism in cutting off the experimental points at the smallest values
of $|t|$ is justified. Both in \cite{PRD106} and in these studies, in all
cases of fitting with allowance for weight matrices, the theoretical curve
always turns out to be systematically shifted relative to the central values
of the experimental points. This effect disappears if the experimental points
with the minimum values of $|t|$ are discarded. Experience shows that for
this it is enough to discard the points for which
$|t|\lesssim 0.005~[\mbox{GeV}^2]$. In the future, we will demonstrate this
effect again without specifying this procedure every time. \vspace{-3.1mm}

\subsection{Separate extraction of parameters from each experiment}
~\vspace{-13.1mm}

\subsubsection{TOTEM}

~\vspace{-7.1mm}

The tables below show the parameters extracted from the fit at different
cut-offs of the data with $|t|$  greater than some
$t_{0~max}$ value specified in the tables. \vspace{-3.1mm}

\begin{figure}[h]
$$\includegraphics[width=130mm]{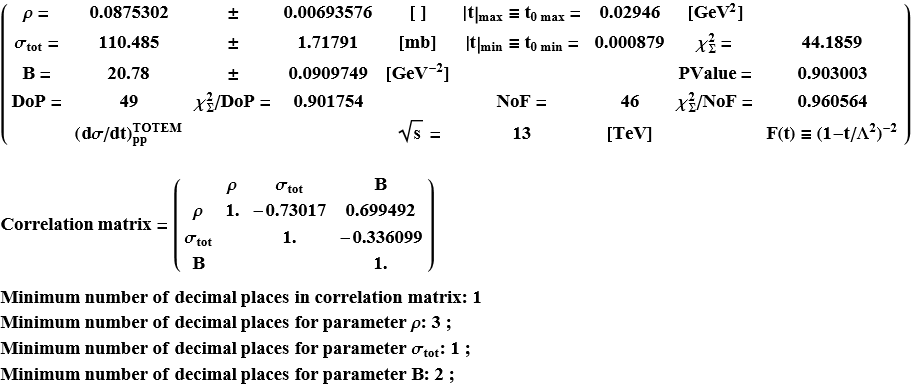}$$
$$\includegraphics[width=130mm]{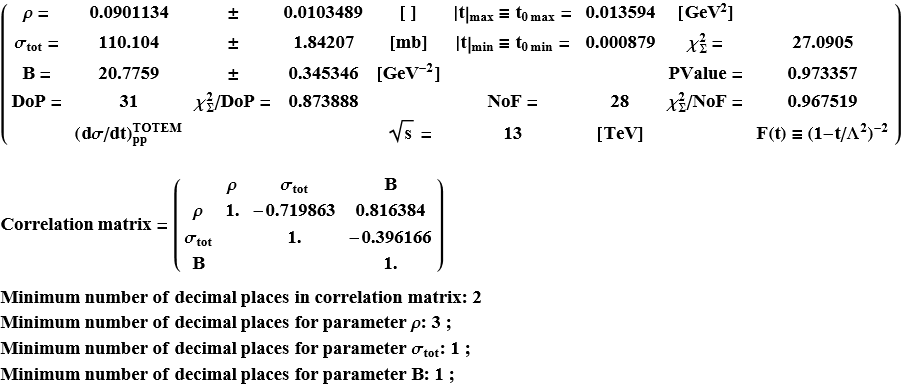}$$
\end{figure}

\begin{figure}[htb]
\noindent
~
\vspace{-3.6mm}
$$
\includegraphics[width=95mm]{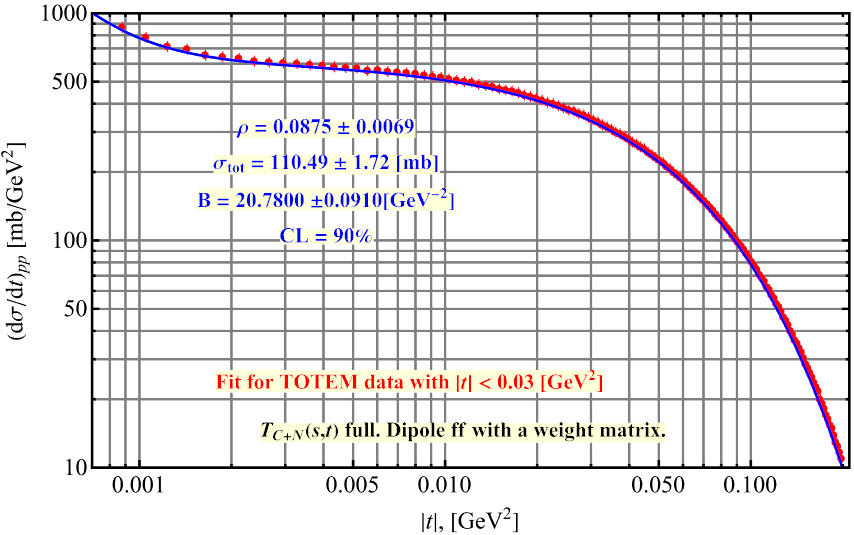}
~
\includegraphics[width=95mm]{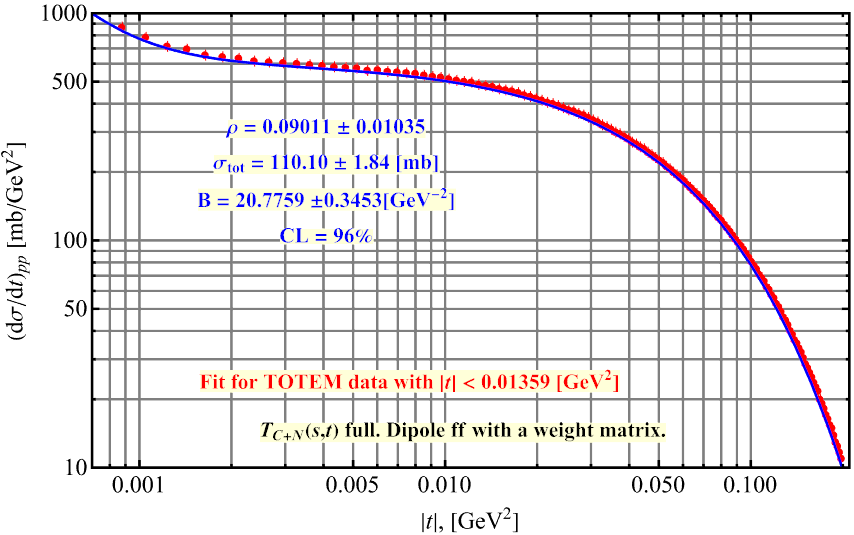}\vspace{-5.1mm}
$$
\caption{{\it Experimental data TOTEM and
theoretical differential cross section curves corresponding to
the above tables.}
\label{Graph1TOTEM}}
\end{figure}

Curve plots are shown in Figs. \ref{Graph1TOTEM}

\begin{figure}[htb]
\vspace{-6.1mm}
$$\includegraphics[width=150mm]{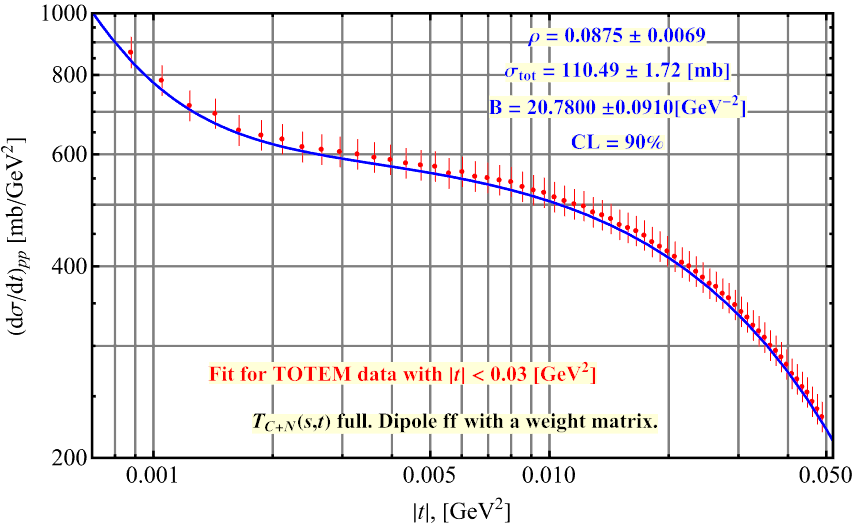}$$
\vspace{-9.1mm}
\caption{{\it Systematic shift of the theoretical curve relative to
experimental points~ -- demonstration of the PPP effect.}}\vspace{-3.6mm}
\label{sdvig}
\end{figure}

On an enlarged scale, it can be seen that the theoretical curves systematically
pass below the experimental points. We have already observed such an effect
in \cite{PRD106}.

This effect is called the PPP effect ("Peele's pertinent puzzle"),
it is associated with the overdetermination (or underdetermination) of systematic
errors and is described in \cite{ppp}. For this reason, the extracted parameters
must be treated  as preliminary estimates. To show this effect, we present
it zoomed in the graph Fig.\ref{sdvig}.

Below we will meet more of such shifts
(without showing zooms).In [1], when cutting off the experimental points
at $|t| > t_{0}$ in the values of the parameters, for sufficiently
small $ t_{0} $ a noticeable instability was observed. For this reason, we
abandoned such experimental points with extremely small $|t|$.
Surprisingly, this made it possible to get rid of this systematic shift while
maintaining a sufficiently high level of confidence.
In this case, we also discarded such points in the above arrays and got the
following results (see tables below):

Graphs corresponding to these tables are shown in 
Fig. \ref{Graph1aTOTEM}

Note that in this case there is now no PPP effect of systematic
shift of the theoretical curve down from the experimental points and a high
level of confidence is retained.
These two plots show two results extracted from experimental data with different
cut-off on the high $ \vert t \vert $ side. Thus, this is a clear demonstration
of the importance of the condition $|t|\rightarrow 0$ to extract the parameter
$\rho$ from the experimental data.
Most likely, one should focus on the values of the parameters from the right
graph in Fig. \ref{Graph1aTOTEM}, because it better corresponds to the definition
of forward observables. As is seen, the parameter $\rho$ significantly changes
its value, while the parameters $\sigma_{\mbox{tot}}$ and $B$ change only
slightly. Of course, in this case, the errors become much larger, Otherwise
it would be necessary to increase the number of points in the fitting interval.

\begin{figure}[htb]
$$\includegraphics[width=139mm]{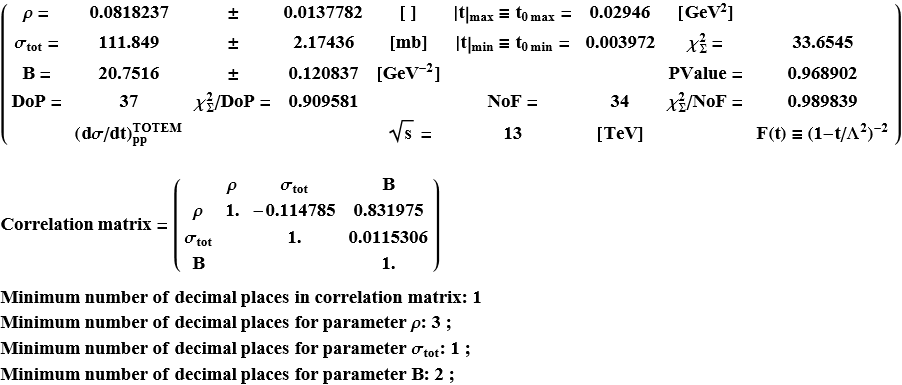}$$
$$\includegraphics[width=139mm]{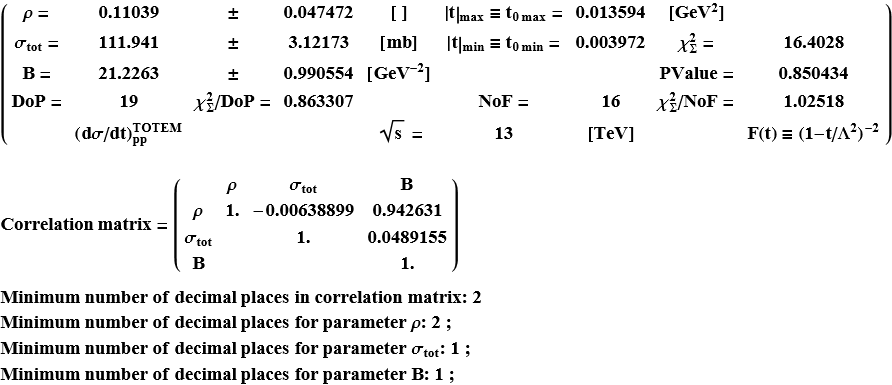}$$~\vspace{-7.1mm}
\label{Table1a-2aTOTEM}
\end{figure}

\begin{figure}[htb]
\noindent
~\vspace{-3.6mm}
$$
\includegraphics[width=95mm]{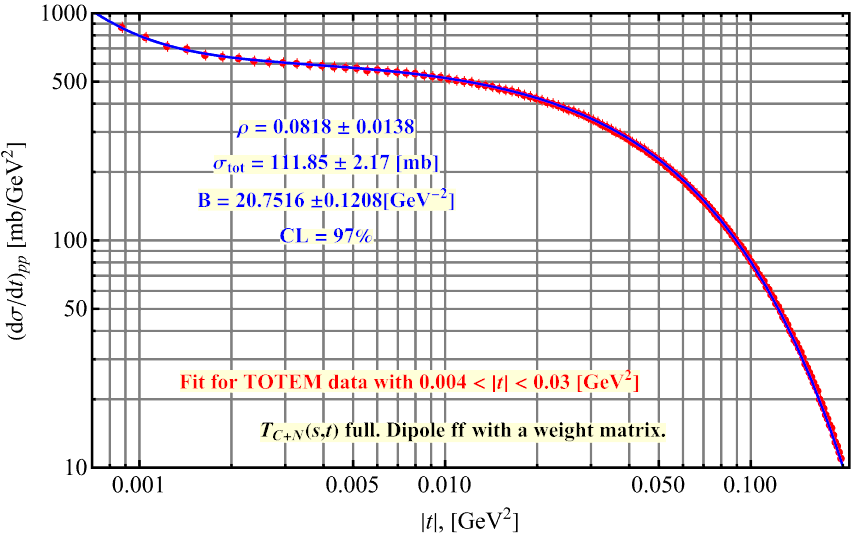}
~
\includegraphics[width=95mm]{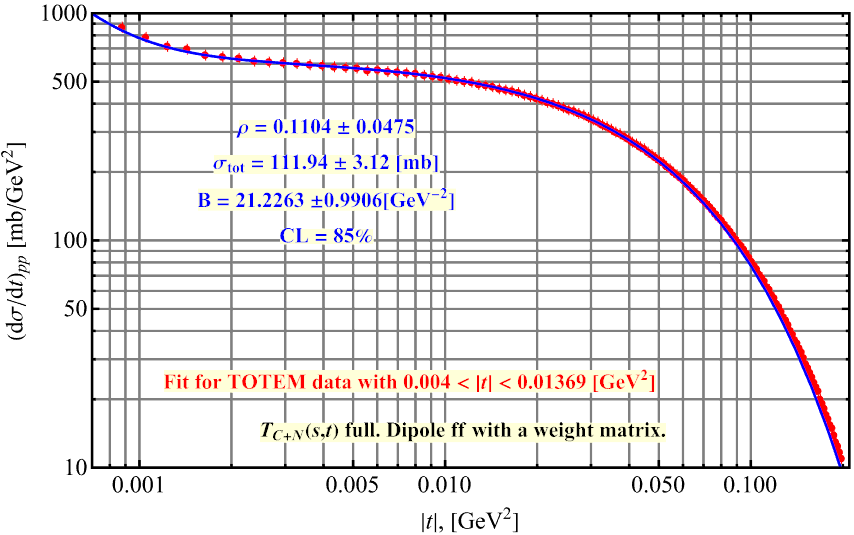}\vspace{-5.1mm}
$$
\caption{{\it Experimental data TOTEM and
theoretical differential cross section curves corresponding to
the above tables.\vspace{-3.1mm}}
\label{Graph1aTOTEM}}
\end{figure}

\newpage

\subsubsection{ATLAS}
~\vspace{-7.6mm}

In this experiment we meet a picture similar to the previous consideration
picture, though in a much tougher version than in
the TOTEM due to the fact that the errors of the ATLAS experiment are much
smaller.

As in the previous subsection, we choose two options for the data cut-off
with large $|t|$ which have higher confidence levels 90{$\%$}: \vspace{-3.6mm}

\begin{figure}[htb]
$$\includegraphics[width=135mm]{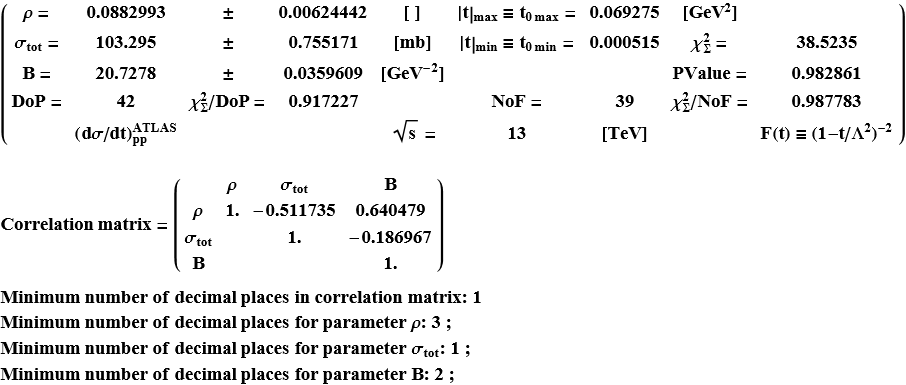}$$
$$\includegraphics[width=135mm]{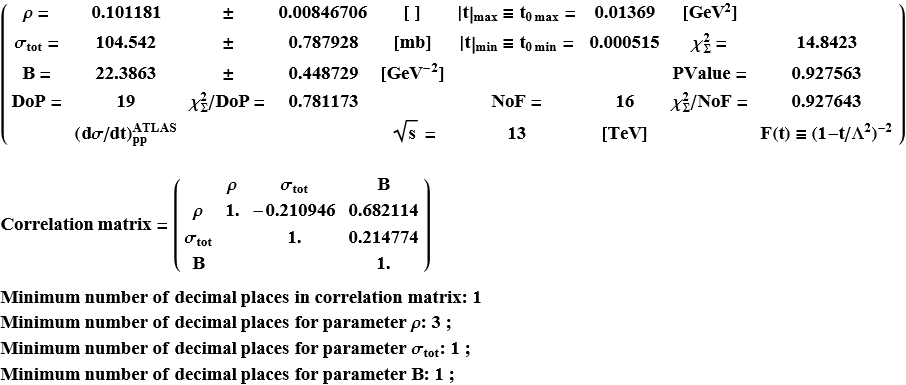}$$~\vspace{-9.1mm}
\label{Table1-2ATLAS}
\end{figure}

The plots corresponding to these tables are shown below in
Fig. \ref{Graph1ATLAS}

\begin{figure}[htb]
\noindent
~\vspace{-3.6mm}
$$
\includegraphics[width=95mm]{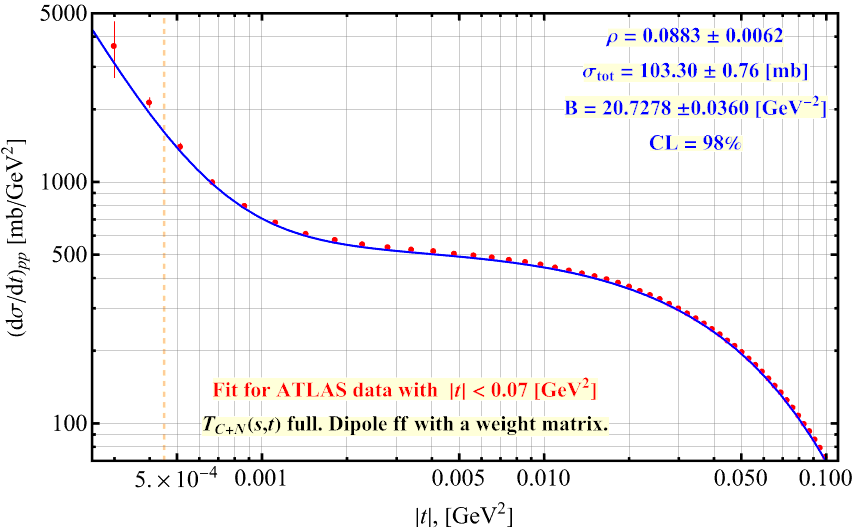}
~
\includegraphics[width=95mm]{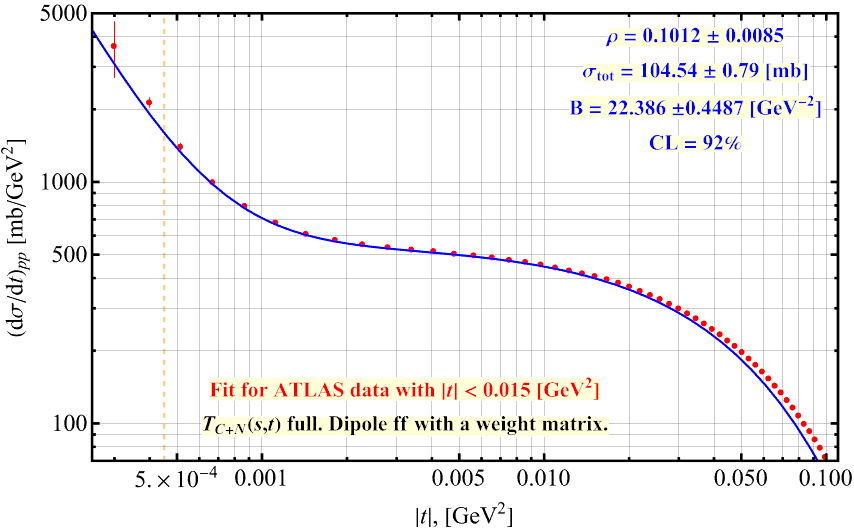}\vspace{-5.1mm}
$$
\caption{{\it Experimental data from ATLAS and
theoretical differential cross section curves corresponding to
the above tables.}}
\label{Graph1ATLAS}
\end{figure}

\newpage

On a larger scale, it is clearly visible that these curves systematically
pass below the central values of the experimental points\footnote{This
effect is called "Peele's pertinent puzzle" ("PPP effect".See the paper
\cite{ppp}). It is a consequence of an overestimation of systematic errors
and manifests itself in the properties of the weight matrix used to compose
the $\chi^2$ function.}, the same as in the case with the TOTEM data described
by us in \cite{PRD106}, although visually this effect not as noticeable as
in the case of TOTEM. So one should take the above values of parameters as
evaluative, not as conclusive ones. However, we note that the values of the
parameters get changed a lot as the values of $|t|$ get closer to $0$,
which is clearly seen from the comparison of parameters in the right plot
of Fig. \ref{Graph1ATLAS} with the left graph in the same figure (or from
previous tables).

The effect of shifting of the theoretical curve, in the case of the TOTEM
data, disappeared, when, for objective reasons, some number of points with
the smallest values of $|t|$ were omitted. In the case of ATLAS 
there are too good reasons to discard experimental points for which
$|t|<0.0025~\mbox{GeV}^2$, as  they lie in a  zone of instability of
parameter values. This leads to the following results:
\begin{figure}[h]
$$\includegraphics[width=140mm]{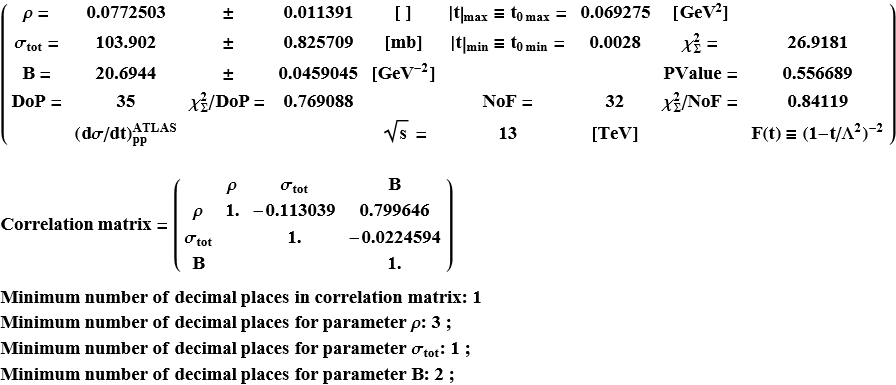}$$
$$\includegraphics[width=140mm]{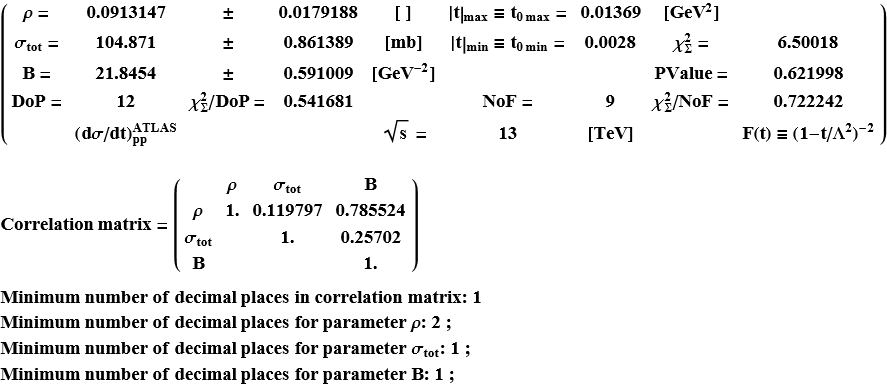}$$~\vspace{-9.1mm}
\label{Table1a-2aATLAS}
\end{figure}

\begin{figure}[htb]
\noindent
~\vspace{-3.6mm}
$$
\includegraphics[width=95mm]{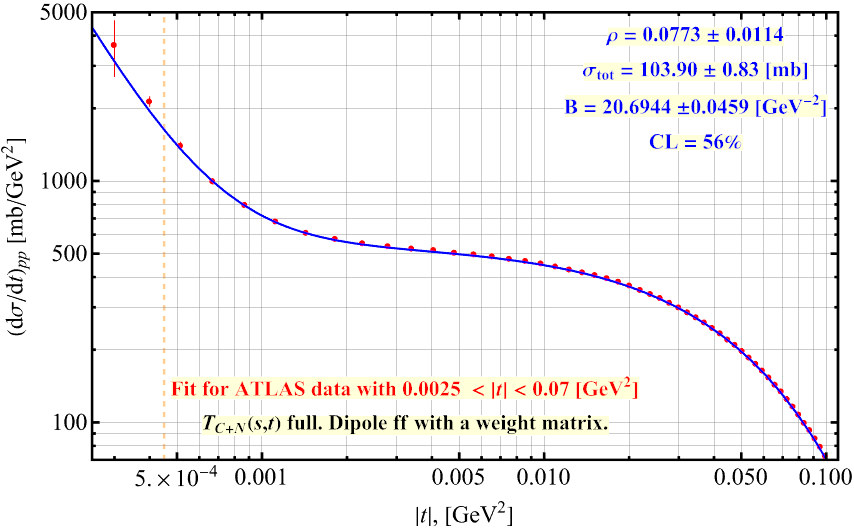}
~
\includegraphics[width=95mm]{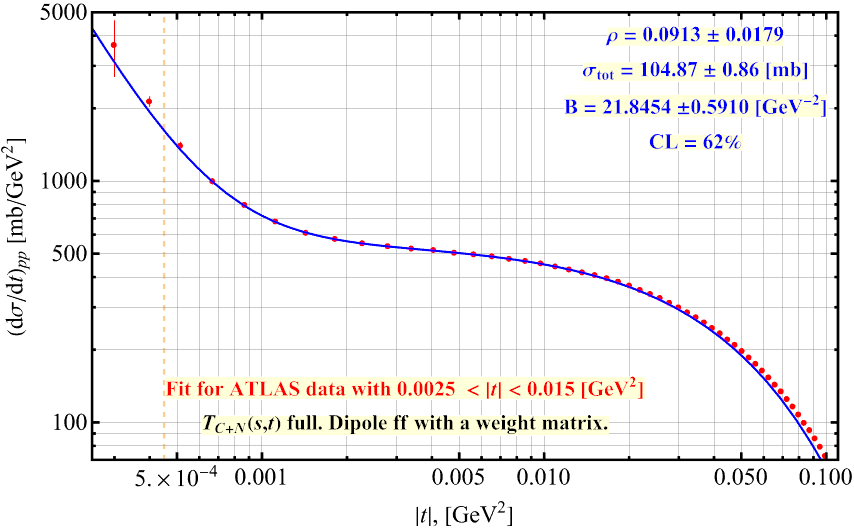}\vspace{-4.1mm}
$$
\caption{{\it Experimental data from ATLAS and
theoretical ATLAS differential cross section curves corresponding to
clipping of experimental data both from above and from below.}}
\label{Graph1aATLAS}
\end{figure}

Curve plots are shown in Figs. \ref{Graph1aATLAS}

As in the case of the TOTEM, in these cases the theoretical curves do not
shifted relative to the central values of the experimental points
(no PPP effect \cite{ppp}) and still show quite a high
confidence level.

There is no reason to reject either of these two options. Certainly
can be inclined in favour of a variant in which experimental
data with smaller values of $|t|$ are held. One cannot be sure, however,
that with the advent of new experimental data  these values will not  also
be unacceptable. Nevertheless, parameter values, from the available experimental
data, shown in the right part of Fig. \ref{Graph1aATLAS} should be considered
the most reliable in the force of what in this case they better correspond
to the definition of the forward ($t\rightarrow 0$) observables to which
all three parameters ($\rho,~ \sigma_{\mbox{tot}},~ B$) belong.
\newpage

~\vspace{-14.1mm}

\section{Conclusions\vspace{-1.6mm}}

\begin{enumerate}
\item In this work, we reanalyzed the data from the TOTEM and ATLAS collaborations.
The result of this processing for the three basic parameters is presented
in the following table:\vspace{2.1mm}

\begin{tabular}{|c|c|c|}\hline
~                         & TOTEM             & ATLAS             \\ \hline
$\rho$                    &$0.1104\pm 0.0475$ &$0.0913\pm 0.0179$ \\
$\sigma_{\mbox{tot}}$ [mb]&$111.94\pm 3.12$   &$104.87\pm 0.86$   \\
$B~\mbox{[GeV}^{-2}]$       &$21.2263\pm 0.9906$&$21.8454\pm 0.5910$\\ \hline
$\mbox{[GeV}^2]$          &$0.004\lesssim |t|\lesssim 0.014$&
$0.003\lesssim |t|\lesssim 0.014$  \\ \hline
\end{tabular}

\begin{itemize}
\item As for the total cross sections, the "Tevatron puzzle" is essentially
reproduced when the extreme values differed (and remained so) by more than
three standard deviations. In our case, the central value of the ATLAS total
cross section  is separated from the central value of the TOTEM total cross
section by more than eight (!) ATLAS standard errors (we take the ATLAS as
a specimen  for the reason that the total errors of this experiment are
fundamentally less than those of the TOTEM experiment). Such a difference
suggests that one should not combine these two arrays of data into a single
one when fitting (we have briefly mentioned this at the beginning of the
article). In addition, we took into account that, in contrast to the full
presentation from the ATLAS, the TOTEM data are not (yet?) placed in HEPDATA.
\item The values of the $\rho$ parameter overlap significantly,taking into
account the error corridors. However, the TOTEM error is huge and amounts
to more than 40{\%}while the ATLAS error is two times smaller, about 20{\%}.
\item As for the parameters $B$,  there is a good agreement, the parameters
are close to each other and do not raise questions.\vspace{-2.6mm}
\end{itemize}
\item Of course, this state of affairs puts, say, the authors of
phenomenological models, in a difficult position. In a way this also
concerns the values of the $\rho$ parameter the accuracies of which
diverge quite noticeably. This does not allow giving preference to any
of its values with an accuracy of no worse than 10{\%} (as the TOTEM
authors claim).
\item The above detailed study of the data from the TOTEM and ALFA/ATLAS
collaborations leads us to the conclusion that a
proper theoretical analysis based on them is impossible without a
fundamental decrease in the difference between
the measured values of the differential cross sections at these
facilities. We emphasize in particular that in both experiments,
the experimental data for which $|t|\lesssim 0.003~\mbox{GeV}^2$
seem doubtful.
\item For a reliable retrieval of CNI
parameters the data in which $|t|\gtrsim 0.07~\mbox{GeV}^2$ are
of no interest.
\item In general, both of these experiments together do not provide
reliable grounds for accurate theoretical estimates of the total
scattering cross section and the $\rho$ parameter, and therefore
cannot yet serve as a solid basis for fundamental physical
conclusions ("Odderon", etc.).
\item The situation similar to that at  the TEVATRON, with
$\sigma_{tot}$ difference between the extreme points  more than
$3\sigma$, Fig. \ref{Fig.9} (left), is not only reproduced  but even
worsened: the difference in $\sigma_{tot}$ values between the two LHC
collaborations at $\sqrt{s}= 13$ TeV is more than $8\sigma_{ATLAS}$.

\begin{figure}[htb]
\noindent
$$
\includegraphics[width=92mm]{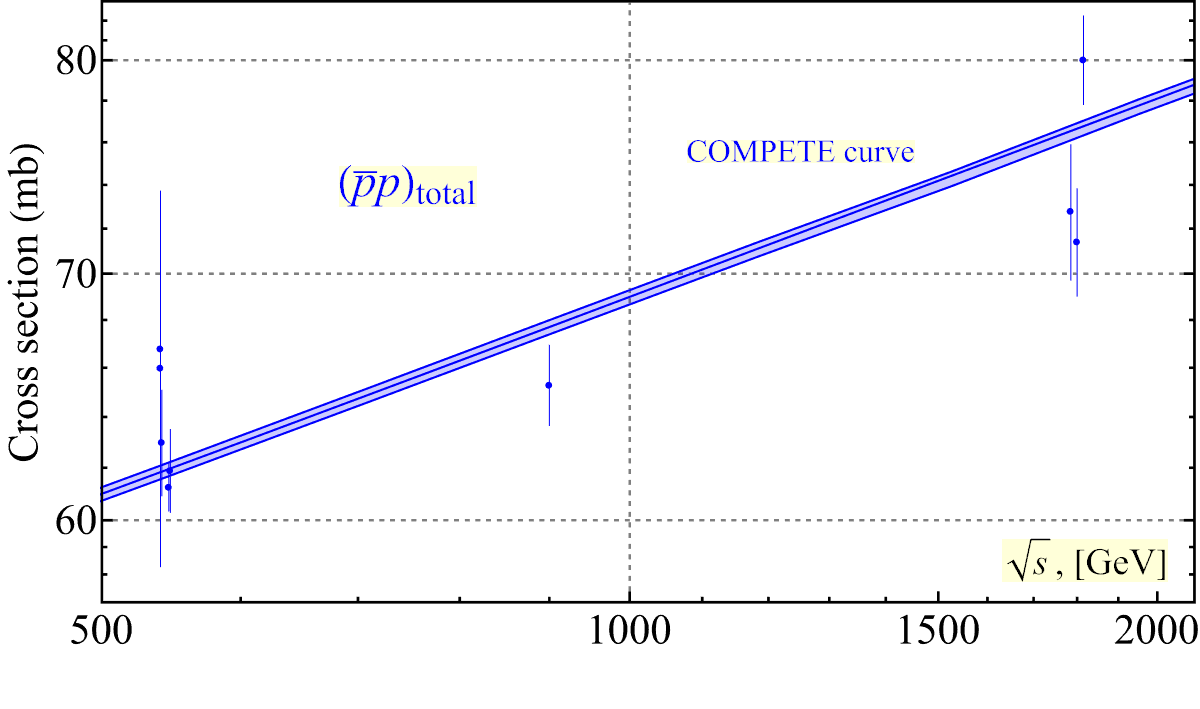}
~
\includegraphics[width=98mm]{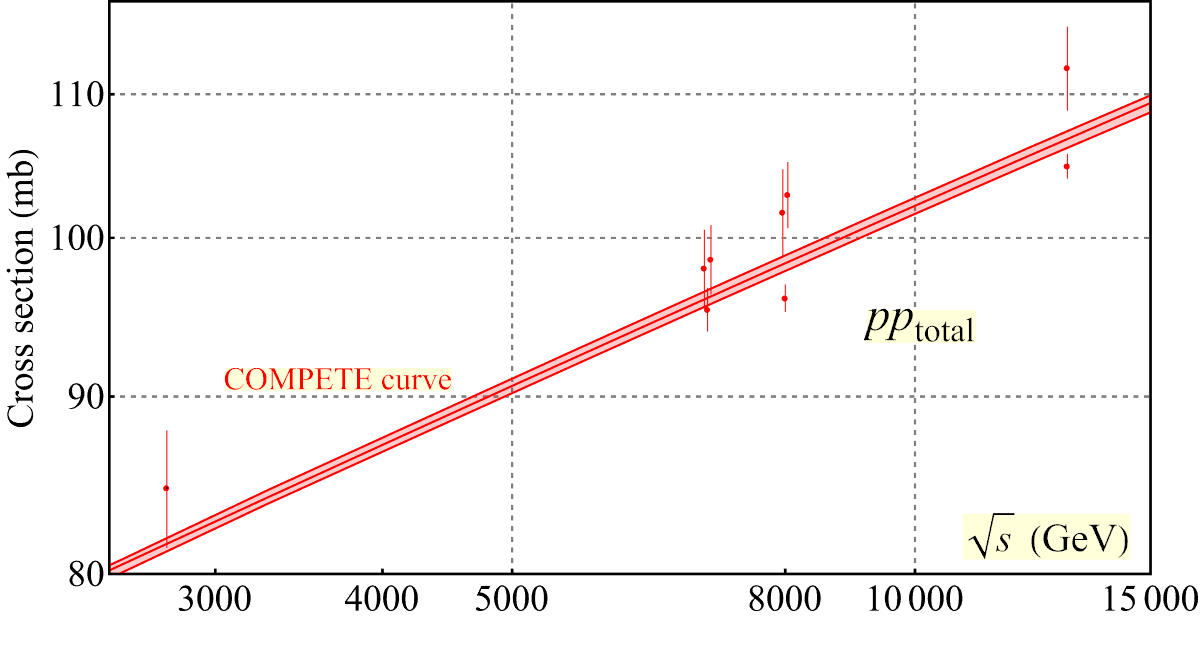}\vspace{-2.1mm}
$$
\caption{
{\it Disagreement  in} $\sigma_{tot}$ {\it at the} TEVATRON {\it at}
$\sqrt{s}= 1.8$ [TeV] ({\it left}) {\it and at the} LHC {\it energies}
({\it right}). {\it For} COMPETE {\it curve see} \cite{COMPETE}.
}
\label{Fig.9}
\end{figure}
\item We would like to emphasize that appreciating the significance of both
experiments very high, in no way  we want to take the pose of lecturing critics
but rather express understandable "consumer" concerns about
danger of a repeat of the "frozen discrepancy" situation  {\it {\`a}~la}
Tevatron. This compels us to address our concerns to our experimental colleagues.
\end{enumerate}
~\vspace{-10.6mm}

\section*{Acknowledgements}
We are grateful to Per Grafstr\"{o}m, Hasko Stenzel, Kenneth
\"{O}sterberg for clarifying correspondence and V.B. Anikeev for stimulating
discussions at early stage of this work.

\end{document}